\documentclass[preprint]{aastex}
\begin{document}

\title{Constraining Dust Extinction Properties via the VVV Survey}

\author{D. Majaess $^{1,2}$, D. Turner $^{2}$, I. D\'{e}k\'{a}ny $^{3}$, D. Minniti $^{4,5,6}$, W. Gieren $^{4,7}$}

\affil{$^1$ Mount Saint Vincent University, Halifax, NS B3M 2J6, Canada.}
\affil{$^2$ Department of Astronomy \& Physics, Saint Mary's University, Halifax, NS B3H 3C3, Canada.}
\affil{$^3$ Astronomisches Rechen-Institut, Zentrum f\"ur Astronomie der Universit\"at
Heidelberg, M\"onchhofstr. 12-14, D-69120 Heidelberg, Germany.}
\affil{$^4$ The Millennium Institute of Astrophysics (MAS), Santiago, Chile.}
\affil{$^5$ Departamento de Ciencias F\'{i}sicas, Universidad Andr\'{e}s Bello, Campus La Casona, Fern\'{a}ndez Concha 700, Santiago, Chile.}
\affil{$^6$ Vatican Observatory, V00120 Vatican City State, Italy.}
\affil{$^7$ Universidad de Concepci\'{o}n, Departamento de Astronomía, Casilla 160-C, Concepci\'{o}n, Chile.}
\email{dmajaess@ap.smu.ca}

\begin{abstract}
Near-infrared color-excess and extinction ratios are essential for establishing the cosmic distance scale and probing the Galaxy, particularly when analyzing targets attenuated by significant dust.  A robust determination of those ratios followed from leveraging new infrared observations from the VVV survey, wherein numerous bulge RR Lyrae and Type II Cepheids were discovered, in addition to $BVJHK_{s}(3.4\rightarrow22)\mu m$ data for classical Cepheids and O-stars occupying the broader Galaxy.   The apparent optical color-excess ratios vary significantly with Galactic longitude ($\ell$), whereas the near-infrared results are comparatively constant with $\ell$ and Galactocentric distance ($\langle E(J-\overline{3.5\mu m})/E(J-K_s) \rangle =1.28\pm0.03$).  The results derived imply that classical Cepheids and O-stars display separate optical trends ($R_{V,BV}$) with $\ell$, which appear to disfavor theories advocating a strict and marked decrease in dust size with increasing Galactocentric distance. The classical Cepheid, Type II Cepheid, and RR Lyrae variables are characterized by $\langle A_{J}/E(J-K_s) \rangle = \langle R_{J,JK_s} \rangle  =1.49\pm0.05$ ($\langle A_{K_s}/A_J \rangle =0.33\pm0.02$), whereas the O-stars are expectedly impacted by emission beyond $3.6 \mu m$.  The mean optical ratios characterizing classical Cepheids and O-stars are approximately $\langle R_{V,BV} \rangle \sim3.1$ and $\langle R_{V,BV} \rangle \sim3.3$, respectively.
\end{abstract}


\section{Introduction}
Applying corrections for dust extinction is a ubiquitous task executed in disciplines throughout astronomy, yet key concerns persist regarding the topic.  Cited color-excess and total-to-selective extinction ratios are contested \citep{tu12,na16}, and a debate continues regarding the compositional nature of dust and the source(s) behind diffuse interstellar absorption lines.  The matter is exacerbated by the dependence of certain extinction ratios ($A_{V}/E(B-V)=R_{V,BV}$) on Galactic longitude $\ell$ \citep{sc16}, as exemplified by observations of the young cluster Westerlund 2 \citep[$\ell \sim 280 \degr$, $R_{V,BV}\sim 4$ versus $\langle R_{V,BV}  \rangle \sim3.1$\footnote{\citet{ti14} obtained $\langle R_{V,BV} \rangle=2.40 \pm 1.05$ from high latitude SDSS BHB stars, whereas \citet{tu76} determined $\langle R_{V,BV} \rangle \sim3.1$ from open clusters.},][]{ca13}.  Yet the challenge inherent in determining those ratios often requires the adoption of results tied to separate sight-lines and stellar populations, with a potential penalty being the propagation of  systematic uncertainties.  Caution is likewise warranted when aiming to subvert such difficulties by assuming a linear relationship between reddening and distance, since numerous sight-lines are characterized by non-linear trends \citep{nk80}.  More broadly, \citet{pk12} and \citet{na16} argued that extinction laws adopted in surveys aiming to constrain cosmological models should be revisited \citep[e.g.,][]{ri11}, and there exist optimal passband combinations displaying less systematic and random scatter \citep[see also][]{ng12}.

As a result of the aforementioned uncertainties, infrared observations are of particular importance when establishing the cosmic distance scale, as the wavelength regime exhibits a reduced sensitivity to dust obscuration relative to optical data (i.e., $A_{\lambda} \sim \lambda^{-\beta}$).  Therefore, potential and often unconstrained variations in the extinction law are less onerous on the uncertainty budget (e.g., $\Delta \mu_0$).  Spitzer observations of Cepheids throughout the Local Group exemplify that advantage \citep{sc11,ma13}, in concert with infrared monitoring of star clusters \citep{ch12,mb13}.  For example, uncertainties associated with the reddening of the Large Magellanic Cloud, a pertinent anchor of the cosmic distance scale, are comparatively marginal in the mid-infrared where $\langle A_{3.6 \mu m}/E(B-V) \rangle \sim 0.18$ \citep{ma13}.    An added advantage of infrared observations is the mitigated impact of compositional differences between calibrating and target standard candles when establishing distances, as line blanketing may affect optical $BV$ observations \citep{cc85,ma09}.  

In this study, reddening and total-to-selective extinction ratios (e.g., $A_{J}/E(J-K_s)$) are inferred from a diverse stellar demographic.  That was accomplished by examining new $JHK_s$ observations from the $VVV$ survey \citep{mi10}, wherein RR Lyrae and Type II Cepheid variables were discovered toward the Galactic bulge and an adjacent region of the Galactic disk, in tandem with multiband infrared (e.g., Spitzer 3.6 $\mu m$) and optical data for O-stars and classical Cepheids throughout the Galaxy.   The analysis aims to provide key insight regarding extinction, and assess whether the data corroborate findings implying a potential link between the Galactocentric distance and dust size, as indicated by red clump stars \citep[][discussion and references therein]{za09,go13}.

\begin{figure}[!t]
\begin{center}
\includegraphics[width=8cm]{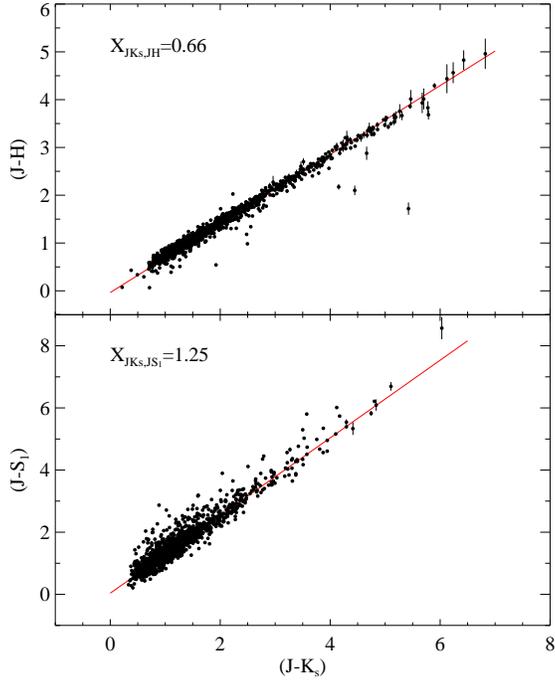} 
\caption{Apparent color ratios for Type II Cepheids (top panel) and RR Lyrae (lower panel) variables identified in the $VVV$ and GLIMPSE surveys ($S_1\simeq 3.6 \mu m$).  The stars are key for establishing a sizable Galactocentric baseline to evaluate reputed extinction ratio variations.  For clarity purposes, only uncertainties along the ordinate are displayed.}
\label{fig-cr}
\end{center}
\end{figure}

\section{Analysis}
\subsection{Determining color-excess \& total extinction ratios}
\label{scolors}
\begin{figure}[!t]
\begin{center}
\includegraphics[width=8cm]{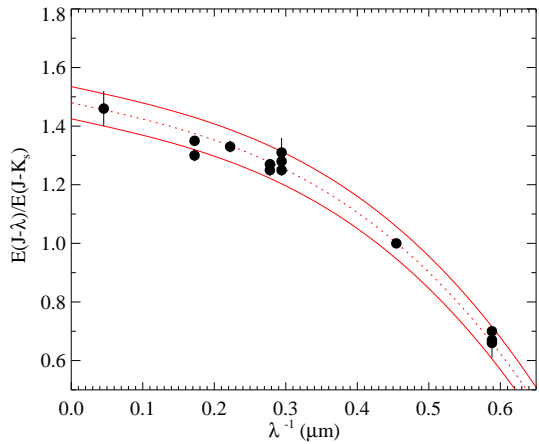} 
\caption{Mean color-excess ratios for the classical Cepheid, Type II Cepheid, and RR Lyrae variables analyzed.  The mean near-infrared total-to-selective extinction ratio determined is $\langle R_{J,JK_s} \rangle =1.49\pm0.05$, and convergence is apparent beyond $3.4 \mu m$.  The bracketing fits represent $\pm 1\sigma$.}
\label{fig-cer}
\end{center}
\end{figure}

The desired extinction ratios can be determined via differing approaches.  A comparison of $E(J-\lambda)/E(J-K_s)$ and $\lambda ^{-1}$ yields $R_{J,JK_s}=A_J/E(J-K_s)$ as $\lambda \rightarrow \infty$, thus providing the coefficient linking the color-excess to the total extinction (Figs.~\ref{fig-cr} \& \ref{fig-cer}). That coefficient may be determined from the apparent stellar colors to bypass potentially uncertain intrinsic colors (e.g., $E(J-K_s)=(J-K_s)-(J-K_s)_0$).   
\begin{eqnarray}
\nonumber
E(J-\lambda)/E(J-K_s) = X_{J \lambda ,JK_s}  \\ \nonumber
\frac{(J-\lambda)-(J-\lambda)_{0}}{(J-K_s)-(J-K_s)_0} = X_{J \lambda ,JK_s} \\ \label{eq1}
(J-\lambda)-(J-\lambda)_{0} = X_{J \lambda ,JK_s} (J-K_s) -X_{J \lambda ,JK_s} (J-K_s)_{0} \\  \label{eq2}
(J-\lambda)= X_{J \lambda ,JK_s}(J-K_s)+\beta
\end{eqnarray}
Where $X$ is the color-excess ratio, and $\beta$ is constant for stars sharing common intrinsic colors. Color-excesses with extended baselines should be selected (e.g., $J-K_s$) to foster diminished uncertainties.  

O-stars are lucrative targets given their small intrinsic color spread \citep[e.g., $(B-V)_0$,][]{tu94}, and are bright along with their longer-period classical Cepheid counterparts. Yet excess infrared emission endemic to O-stars and their ambient environment may bias mid-infrared colors (e.g., $W_4 \sim 22 \mu m$), and hence determinations of the desired extinction ratios.  The angular resolution associated with longer-wavelength mid-infrared photometry is likewise too coarse for dense nascent regions, although Spitzer observations are preferred owing to their improved resolution relative to WISE.  Indeed, the $W_2$ ($4.6 \mu m$) reddening ratio cited in Table~\ref{table1} for O-stars is systematically larger than inferred from the bracketing Spitzer photometry ($S_2,S_3$), which hints at the onset of contamination.  

Intrinsic colors for RR Lyrae and Cepheid variables are linked to the pulsation period (i.e., $M_{\lambda_1}-M_{\lambda_2} \sim \alpha \log{P} + \beta$). However, the expression is an approximation, especially for optical passbands where the instability strip exhibits sizable width (i.e., a temperature spread) and a metallicity term may be significant.  Rotation may likewise play a role \citep{an14}.  Therefore, variables with similar pulsation periods may feature different absolute optical magnitudes.  Yet at infrared wavelengths the effect is relatively marginal since the observations primarily track radius changes, rather than those tied to the temperature.  For certain variable classes, Eqn.~\ref{eq1} may consequently be recast as:
\begin{eqnarray}
\nonumber
(J-\lambda)&=& X_{J \lambda ,JK_s} (J-K_s)-X_{J \lambda ,JK_s}(J-K_s)_{0} + (J - \lambda)_{0} \\ \nonumber
(J-\lambda)&=& X_{J \lambda ,JK_s} (J-K_s)-X_{J \lambda ,JK_s}(\alpha \log{P} +\beta)+(\gamma \log{P} +\delta) \\ \nonumber
(J-\lambda)&=& X_{J \lambda ,JK_s} (J-K_s) + \log{P} (-X_{J \lambda ,JK_s} \alpha + \gamma) -x \beta +\delta \\ \label{eq5}
(J-\lambda)&=& X_{J \lambda ,JK_s} (J-K_s)+\zeta \log{P} + \eta
\end{eqnarray}
A caveat to Eqn.~\ref{eq5} arises when treating $4.5 \mu m$ and $4.6 \mu m$ Cepheid photometry, as the period-color relation becomes non-linear.  The $(3.6 \mu m - 4.5 \mu m)$ color is approximately constant for shorter-period Cepheids, and subsequently transitions into a bluer convex trough at longer periods \citep{ma13}.  That trend is partly attributable to the temperature dependence of CO absorption and dissociation \citep{hg74,sc11}.  RR Lyrae variables are significantly hotter than classical or Type II Cepheids, and likely immune to that effect. The pulsation term may be negligible for infrared passbands owing to the aforementioned reduced temperature dependence (Rayleigh-Jeans tail of the Planck function).   

The following total extinction ratio may now be evaluated from the total-to-selective extinction ratio derived via Eqn.~\ref{eq2} (or \ref{eq5}).
\begin{eqnarray}
\label{eq3}
A_{K_s}/A_{J}&=&1-{R_{J,JK_s}}^{-1}
\end{eqnarray}
Separate total extinction ratios ($A_{\lambda_1}/A_{\lambda_2}$) may be established once one is known (Eqn.~\ref{eq3}), and by making use of the expression described by Eqns.~\ref{eq2} or \ref{eq5}.
\begin{eqnarray}
\label{eq4}
A_\lambda / A_J &=&  (-1/R_{J,JK_s})\frac{(J-\lambda)-\beta}{(J-K_s)}+1 \\ \nonumber
&=& ( A_{K_S}/A_J-1) \frac{E(J-\lambda)}{E(J-K_{s})}+1  \\ \nonumber
&=&\frac{A_J A_{K_s}-A_{\lambda} A_{K_s} -A_{J}^2+A_{\lambda} A_{J} + A_{J}^2-A_J A_{K_s}}{A_J(A_J-A_{K_s})} \\ \nonumber
&=&A_\lambda / A_J
\end{eqnarray}
The final method employed here to assess the total-to-selective extinction ratio involves exploiting a sample at a common distance.  RR Lyrae and Cepheids present an advantage since their intrinsic color and magnitude may be estimated from their pulsation period (i.e., $M_{\lambda} \sim \alpha \log{P}+\beta$), and contamination by AGB and red giant stars is less problematic.  A disadvantage is that those variables are not as ubiquitous as red clump stars.  
\begin{eqnarray}
\nonumber
K_s - M_{K_s}-A_{K_s}=\mu_0 \\ \nonumber
K_s-(\alpha \log{P} +\beta) - R_{K_s,JK_s} ((J-K_s)-(J-K_s)_0)=\mu_0  \\ \nonumber
K_s-\alpha \log{P} - \beta - R_{K_s,JK_s} (J-K_s)-R_{K_s,JK_s}(\gamma \log{P}+\delta)=\mu_0  \\ \nonumber
K_s=R_{K_s,JK_s} (J-K_s)+ \zeta \log{P}+\eta \\ \label{eq7}
\end{eqnarray}
Sizable samples can be binned as a function of the pulsation period (e.g., bulge RR Lyrae variables), and consequently Eqn.~\ref{eq7} may be transformed to:
\begin{equation}
\label{eq8}
K_s=R_{K_s,JK_s} (J-K_s)+ \tau
\end{equation}
The common distance approach warrants caution when applied to targets along certain bulge sight-lines, where significant substructure, a rich stellar field, and sizable extinction prevail.  \citet{sm07} note that stellar distances along such sight-lines can be biased by blending, and linked parameters would be compromised \citep[see also][]{ma10}.

\begin{figure}[!t]
\begin{center}
\includegraphics[width=8cm]{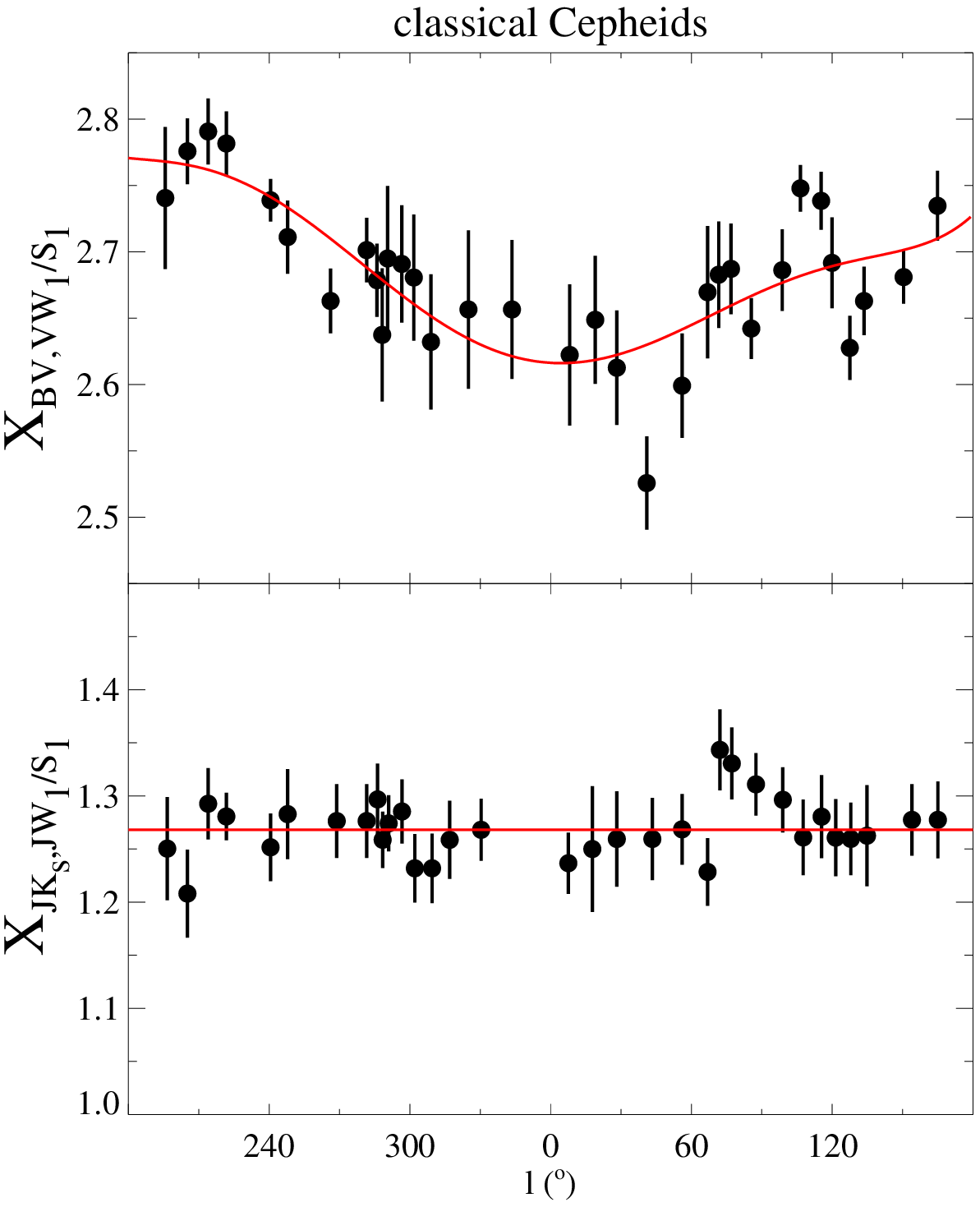}  
\includegraphics[width=8cm]{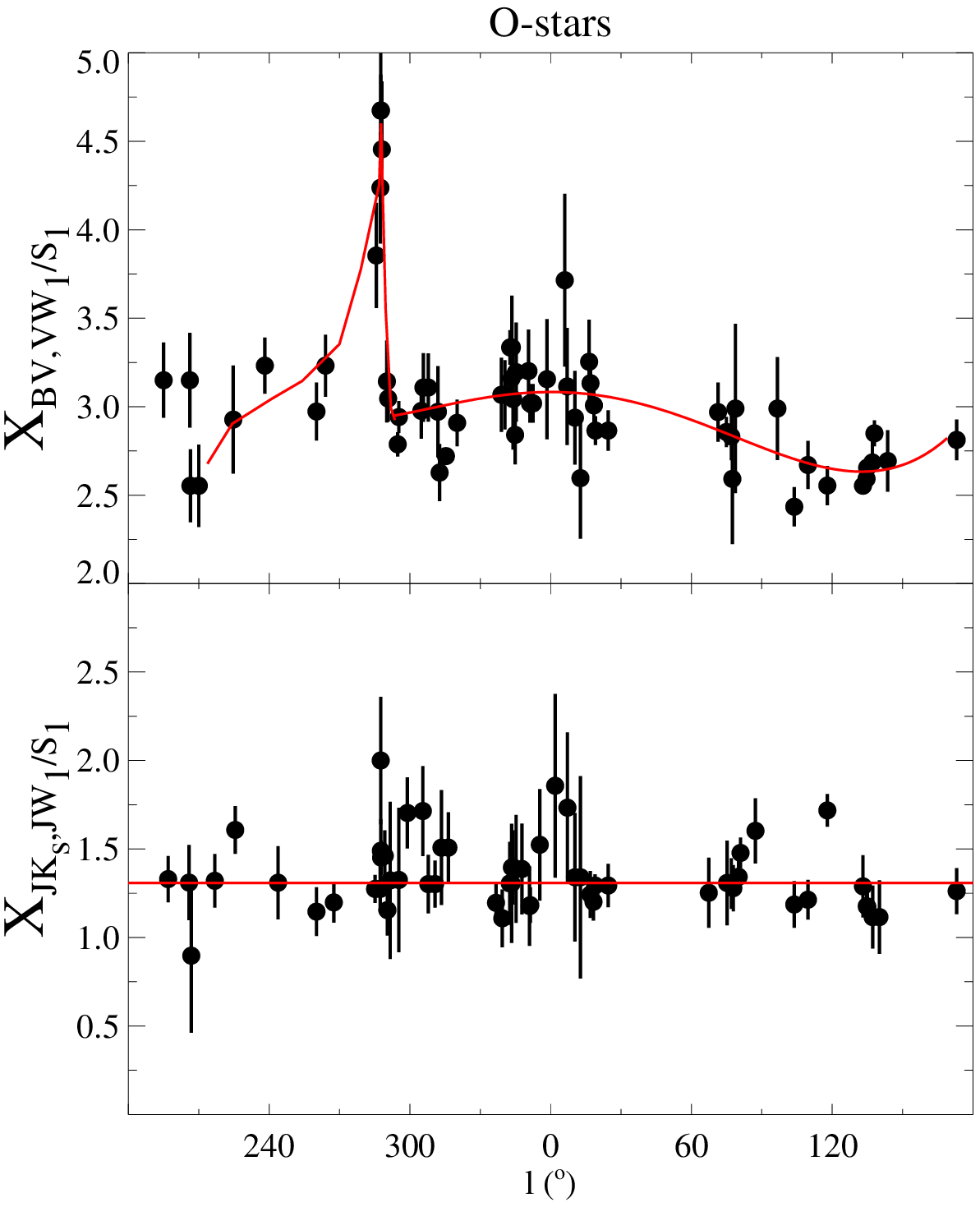} 
\caption{Optical color-excess ratios vary as a function of the Galactic sight-line (top panels), in relative contrast to those tied solely to infrared photometry (e.g., $E(J-\overline{3.5\mu m})/E(J-K_s)$, bottom panels).  Classical Cepheids and O-stars display different optical trends with extrema along separate sight-lines.  The binned results were overlaid with arbitrary (running mean and polynomial) fits to highlight the variations.}
\label{fig-cel}
\end{center}
\end{figure}

\subsection{O-stars and the Variable Star Demographic}
\label{s-unc}
\citet{ma13} utilized O-stars to ascertain an approximate relation between the optical and mid-infrared color-excess and total-to-selective extinction ratios (i.e., $E(3.6 \mu m - 4.5 \mu m)/E(B-V)$ and $A_{3.6 \mu m}/E(B-V)$).  That analysis is now expanded to the broader Galaxy and other stellar populations, namely $\sim 300$ O-stars and classical Cepheids, and $\sim 1.5 \times 10^3$ Type II Cepheids and RR Lyrae variables.

Data from release v1.1 of the Galactic O-star spectroscopic survey \citep{ma04,so11,so14} were correlated with 2MASS, GLIMPSE, and WISE observations \citep[e.g.,][]{wr10}.  The CDS X-Match Service was employed to match the data, typically using $r<1.5\arcsec$.  Apparent colors $(J-\lambda)$ were compared to $(J-K_s)$, where $\lambda=JHK_sW_1S_1S_2W_2S_3$ (see Table~\ref{table1}).  The $S_4$ ($8 \mu m$) and $W_3$ ($12 \mu m$) passbands were ignored to circumvent a significant absorption profile putatively associated with silicate. 

The classical Cepheid analysis was carried out using optical observations compiled by \citet{be00}, in concert with infrared 2MASS, GLIMPSE, and WISE data.  Increased statistics were favored over a limited sample tied to multi-epoch mean magnitudes.  The $4.5 \mu m$, $4.6 \mu m$, and $12 \mu m$ data were ignored when determining the total-to-selective extinction ratio for reasons described above (i.e., CO and the broad silicate absorption profile).  Conversely, the baseline was extended to $W_4$ ($22 \mu m$) as contributions from dust (re)emission appear benign relative to the situation for O-stars.    

A search was undertaken for RR Lyrae variables and Type II Cepheids in the DRV4 release of the $VVV$ survey. The observations provide a critical extension of the Galactocentric baseline, thereby permitting key trends to be identified.  The $VVV$ survey is a near-infrared, wide-field, multi-epoch, and high-resolution campaign monitoring stars in the Galactic bulge, and an adjacent region of the disk  \citep{mi10,sa12}. The survey, which was carried out from the VISTA 4-m telescope (Paranal), strived to extend existing near-infrared surveys beyond their faint limit by $\sim3-6$ magnitudes.  Preliminary analyses of the wide-field campaign, which covers nearly 520 square degrees, indicate that in excess of a billion celestial targets were detected.  To identify the variable stars sought, lightcurve (Fourier) templates were inferred from RR Lyrae and Type II Cepheids discovered by the OGLE survey \citep{so11a,so11b}, and observed by the $VVV$ survey.  The templates were then used to identify variables throughout the broader region where the surveys lack overlap.  The GLS2 period-search algorithm was employed \citep{zk09}, and detections were screened against the templates via a $\chi^2$ scheme.   Type II Cepheids and classical Cepheids display similar lightcurves, and a clean separation of the populations amid photometric uncertainties is challenging.  However, the two populations share similar colors, making their differentiation in this instance relatively unimportant as the broader trends are sought, and the population II class dominates within the bulge \citep{de15}.  A broader discussion concerning the topic is deferred to Hajdu et al. 2016 (in preparation).  Aperture photometry for the RR Lyrae and Type II Cepheid variables were adopted verbatim from the CASU pipeline \citep[][]{mi10}, which include photometric zero-points tied to extinction estimates.  Those extinction corrections are potentially sizable for shorter wavelength VVV Z and Y photometry, which are not investigated here, and are comparatively small for J, H, and $K_s$ bands. The study's broader conclusions are relatively unaffected owing to such small corrections being incorporated into color ratios.  The CASU data were subsequently tied to 2MASS using secondary standards, and the transformation coefficients were near unity as the filter sets are similar.  The principle (systematic) uncertainty is likely attributable to the nature of the aperture photometry utilized (\S \ref{s-res}).  Certain brighter Type II Cepheids are saturated in the $VVV$ survey, and possess inaccurate photometry.  Indeed, saturation occurs for the longer-period and least obscured Type II Cepheids throughout the bulge, and variables at the structure's forefront.  A final catalog of variables detected in the $VVV$ survey will be published in a separate study, where a rigorous discussion of completeness is provided (e.g., accounting for biases toward obscured regions and areas with limited multi-epoch sampling, D{\'e}k{\'a}ny et al. 2016, in preparation).

A $\sigma$-clip (generally 3$\sigma$) approach was adopted to mitigate spurious data, in concert with fitting algorithms that avert outliers (e.g., robust fitting).  Photometric color uncertainties from separate surveys are inhomogeneous, and may bias fits toward brighter and less-reddened objects.  Indeed, the apparent color ratios of the latter stars exhibit sizable uncertainties, and may skew results toward higher ratios.  Thus establishing a broad color baseline assumed priority, and sizable statistics ensured that the formal random uncertainties were often dominated by systematics linked to the fitting algorithms.   Uncertainties cited in Table~\ref{table1} are tied to the spread in the latter.  
 
\begin{figure}[!t]
\begin{center}
\includegraphics[width=8cm]{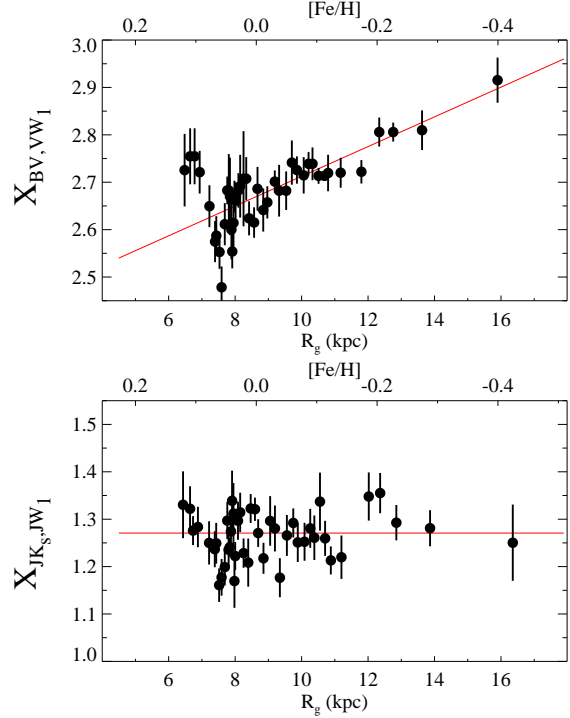} 
\caption{Optical color-excess ratios (binned) inferred from classical Cepheids exhibit a dependence on the \textit{pseudo} Galactocentric distance or metallicity (likely largely stellar in origin).  That contrasts the results tied solely to the infrared photometry analyzed (e.g., right panel, $E(J-3.4 \mu m)/E(J-K_s)$).  In concert with the trend delineated by O-stars (Fig.~\ref{fig-cel}), the findings contradict a paradigm linking a strict and marked decreasing dust grain size with increasing Galactocentric distance.  The \citet{lu11} metallicity gradient is conveyed.}
\label{fig-ceg}
\end{center}
\end{figure}

\begin{table}[!t]
\begin{center}
\small
\caption{Color-excess ratios. \label{table1}}
\begin{tabular}{ccc}
\hline \hline
$\langle E(J-\lambda)/E(J-K_s) \rangle$ & $\lambda$ ($\mu m$) & $Demographic$   \\
\hline
0 &		J ($1.2$)	&		--- \\
$0.66\pm0.05$	&	H	($1.7$)&	Type II Cepheids	\\
$0.67\pm0.01$	&	H	($1.7$)&	RR Lyrae	\\
$0.70\pm0.01$	&	H	($1.7$)&	classical Cepheids	\\
$0.63\pm0.01$	&	H	($1.7$)&	O-stars	\\
1		&	$K_s$ ($2.2$)	&	---	\\
$1.28\pm0.04$	&	$W_1$ ($3.4$)	&	classical Cepheids	\\
$1.25\pm0.02$	&	$W_1$ ($3.4$)	&	Type II Cepheids	\\
$1.31\pm0.05$	&	$W_1$ ($3.4$)	&	RR Lyrae	\\
$1.33\pm0.02$	&	$W_1$  ($3.4$)	&	O-stars	\\
$1.27\pm0.01$	&	$S_1$	($3.6$) &	Type II Cepheids	\\
$1.25\pm0.02$	&	$S_1$ ($3.6$) &	RR Lyrae	\\
$1.28\pm0.01$	&	$S_1$  ($3.6$)	&	O-stars	\\
$1.33\pm0.02$	&	$S_2$	($4.5$)&	RR Lyrae	\\
$1.37\pm0.09$	&	$S_2$	($4.5$)&	O-stars	\\
$*1.54\pm0.01$	&	$W_2$	($4.6$)&	O-stars	\\
$1.35\pm0.01$	&	$S_3$	($5.8$)&	Type II Cepheids	\\
$1.30\pm0.02$	&	$S_3$	($5.8$)&	RR Lyrae	\\
$1.48\pm	0.06$	&	$S_3$	($5.8$)&	O-stars	\\
$1.46\pm0.06$	&	$W_4$ ($22$)	&	classical Cepheids	\\
\hline 
$\langle E(V-\lambda)/E(B-V) \rangle$ &  &  \\
\hline
$2.26\pm0.03	$	&	J ($1.2$) &	O-stars	\\
$2.24\pm0.09	$	&	J ($1.2$)	 &	classical Cepheids	\\
$2.58\pm0.04	$	&	H ($1.7$)	&	O-stars	\\
$2.54\pm0.06	$	&	H ($1.7$)	&	classical Cepheids	\\
$2.77\pm0.06	$	&	$K_s$ ($2.2$)	&	O-stars	\\
$2.70\pm0.06$	&	$K_s$ ($2.2$)	&	classical Cepheids	\\
$2.94\pm0.06	$	&	$W_1$ ($3.4$)	&	O-stars	\\
$2.74\pm0.02	$	&	$W_1$ ($3.4$)	&	classical Cepheids	\\
$3.04\pm0.05	$	&	$S_1$ ($3.6$)	&	O-stars	\\
$3.13\pm0.01	$	&	$S_2$ ($4.5$)	&	O-stars	\\
$2.98\pm0.04$	&	$W_2$ ($4.6$)	&	O-stars	\\
$3.24\pm0.06$	&	$S_3$ ($5.8$)	&	O-stars	\\
$3.10\pm0.26	$	&	$W_4$ ($22$)	&	classical Cepheids	\\
\hline
\end{tabular} \\  
\end{center}
\textit{Notes}: Near-infrared observations are on the 2MASS system, and $S_x$ and $W_x$ signify Spitzer and WISE data, respectively.  Contamination from the environment encircling O-stars is probable beyond $J-5.8 \mu m$, and the lower-resolution $J-4.6 \mu m$ ($W_2$) passband relative to the bracketing Spitzer filters ($S_1$,$S_2$).  The uncertainties cited reflect the spread between the fitting algorithms (see \S \ref{s-unc}).  
\end{table}

\section{Results}
\label{s-res}
The results stemming from infrared photometry of such diverse populations point toward certain similar color-excess ratios (Table~\ref{table1}), although the O-star ratios digress rapidly beyond $3.6 \mu m$, likely owing to infrared excess and contamination from the ambient young environment surrounding the targets.  Bandwidth effects may likewise offset the ratios determined \citep[for a discussion on divergent interpretations see][]{be96}.  Bandpasses are not infinitesimally small, and hence differential extinction presumably occurs across the filter.  The impact partly depends on the stellar energy distribution and dust attenuation. 

Optical ratios ($E(V-\lambda)/E(B-V)$) diverge significantly as a function of $\ell$, and the bulk trends differ between the classical Cepheid and O-star populations (Fig.~\ref{fig-cel}).  Conversely, relatively slight deviations of $E(J-\overline{3.5 \mu m})/E(J-K_s)$ are observed as a function of $\ell$.  The results were subsequently binned to reduce uncertainties.  Furthermore, the residuals tied to $A_J/E(J-K_s)$ are not readily correlated with a \textit{pseudo}\footnote{The result conveyed in Fig.~\ref{fig-ceg} is tied to the integrated extinction properties along the Sun-target sight-line, and is merely an approximation of any Galactocentric dependence.} determination of the Galactocentric distance (Figs.~\ref{fig-cel} \& \ref{fig-ceg}).   The Galactocentric distance was computed according to the following expression, $R_G=\sqrt{d^2 \cos{b}-2 R_0 d \cos{b} \cos{\ell}+{R_0}^2}$, where $R_0$ is the distance to the Galactic centre and $d$ is the classical Cepheid distance.  The distance was evaluated after examining new near-infrared observations of LMC classical Cepheids by \citet{ma15}, from which the following first-order absolute period-Wesenheit relation was established: $W_{J,JK_s}-W_{0,J,JK_s}=J-R_{J,JK_s} (J-K_s)-(-3.18 \log{P}-2.67)=\mu_0$ \citep[assuming $\mu_{0,LMC}=18.45$, see also][]{gi15}.  That Wesenheit function \citep[reddening-free,][]{ma82} was adopted since $R_{J,JK_s}$ appears relatively constant across the Galaxy to within the uncertainties, thus mitigating uncertainties associated with color-excess estimates.  A similar exercise was not carried out for the O-stars granted their absolute magnitudes are acutely sensitive to the spectral classifications assigned, in stark contrast to their intrinsic colors \citep{tu94}.   

\begin{figure}[!t]
\begin{center}
\includegraphics[width=8.5cm]{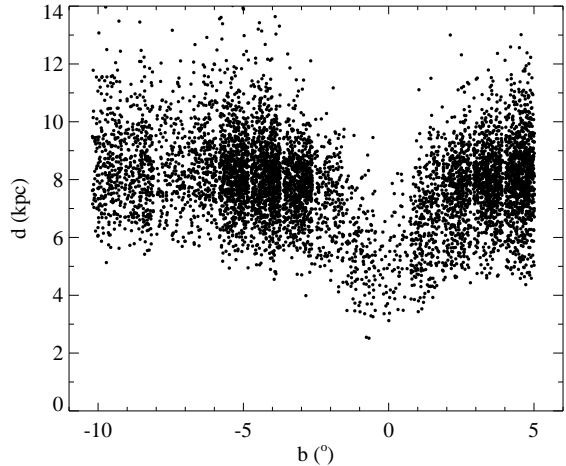} 
\caption{Caution is warranted when applying the common distance approach (Eqn.~\ref{eq8}) to infer extinction properties for stars along the Galactic center sight-line, where extinction and blending are acute ($\ell, b \rightarrow 0 \degr$).  Proximate distances observed for bulge RR Lyrae are likely linked to flux introduced by neighboring stars.}
\label{fig-rr}
\end{center}
\end{figure}

The mean near-infrared total-to-selective extinction ratio for classical Cepheids, Type II Cepheids, and RR Lyrae variables is $\langle A_{J}/E(J-K_s) \rangle=\langle R_{J,JK_s} \rangle=1.49\pm0.05$, which implies $\langle A_{K_s}/A_J \rangle=0.33\pm0.02$ via Eqn.~\ref{eq3}.  The total-to-selective extinction ratio was derived from a polynomial fit with preferential weighting towards the longest-wavelength passband.  Sight-lines ($\ell$) projected along the Galactic bulge are not characterized by \textit{highly} anomalous near-infrared color-excess ratios, while optical data for O-stars adhere to larger ratios relative to the mean (ultimately, multiepoch and multiband PSF photometry with increased sample resolution is preferable for the bulge).  However, the common distance approach described by Eqns.~\ref{eq7} and \ref{eq8} yielded a discrepant finding relative to the apparent color approach, namely that $R_{J,JK_s}\sim1$ characterizes the bulge RR Lyrae variables.  The former requires strict constraints on the sample analyzed, and bulge substructure further complicates the analysis.   \citet{ma10} noted there is a bias for smaller $R_0$ because stars are preferentially sampled toward the near side of the bulge owing to extinction, and an uncertainty in characterizing how a mean distance to the group relates to $R_0$.  Furthermore, it was remarked that values of $R_0$ may be biased by blending, and certain determinations of total-to-selective extinction ratios for bulge sight-lines are impacted.  The density and surface brightness increase markedly as $\ell, b \rightarrow 0 \degr$.  Evidence of blending emerges when assessing the computed RR Lyrae distances as a function of $|b|$ (Fig.~\ref{fig-rr}).  Distance offsets become readily apparent for variables near the Galactic centre, as RR Lyrae become brighter because of contaminating flux from neighboring stars.  The aperture photometry utilized is particularly sensitive to that effect.  A similar trend is present for RR Lyrae variables near the bright and crowded centres of globular clusters \citep{ma12a,ma12b}.  The value of $R$ required to resolve the problem is inconsistent with the results implied by the stellar colors.  The distances were computed using $K_s$ and $J$ period-magnitude relations inferred from LMC and nearby RR Lyrae stars \citep[e.g.,][]{bo09}. 

The optical reddening and total-to-selective extinction ratios correlate with $\ell$.  For the O-stars examined, a maximum $R_{V,BV}$ is observed along the $\ell \sim 290 \degr$ sight-line, and minima are located toward $\ell \sim 133 \degr$ and $\ell \sim 78 \degr$.  A separate trend is delineated by the classical Cepheid optical data, where extrema lie near the anti-centre and $\ell \sim 30 \degr$.  The O-star results are inconsistent with a strict Galactocentric dependence, as the $R_{V,BV}$ extrema are offset by $\sim 180\degr$ ($\ell \sim 78 \rightarrow 280 \degr$), and a higher ratio characterizes the Galactic bulge sight-line. Conversely, the optical ratio characterizing classical Cepheids features a Galactocentric metallicity or distance dependence (Fig.~\ref{fig-ceg}).  That Cepheid trend may be partially explained by metallicity-dependent optical stellar (Cepheid) colors.  \citet{ma09} identified that a $BV$ period-magnitude relation tied to Galactic classical Cepheids yields distances featuring a period-dependence when applied to lower-abundance LMC Cepheids.  \citet{cc85} likewise suggested that the $BV$ color is sensitive to metallicity, whereby lower-abundance Cepheids exhibit bluer intrinsic colors.  Therefore the larger optical color-excess ratios observed toward the anti-centre are expected, as the Galactocentric metallicity gradient shifts to lower abundances.  Hitherto the results disfavor the interpretation that classical Cepheid and O-star optical colors convey that markedly smaller grains dominate at increasing Galactocentric distance, as indicated by certain red clump analyses.   Dust extinction is a complex phenomenon sensitive to a suite of variables (size, density, chemical composition, environment, etc.).  Admittedly, if the Cepheid metallicity ($BV$) effect described is significantly larger than model predictions then the trend in Fig.~\ref{fig-ceg} could potentially invert, and thus support red clump predictions. Yet, the infrared-based Cepheid results would remain unchanged, and in concert with the O-star findings are irreconcilable with that theory. The mean optical total-to-reddening ratios are $\langle A_{V}/E(B-V) \rangle= \langle R_{V,BV} \rangle \sim3.1$ (classical Cepheids) and $\langle R_{V,BV} \rangle\sim3.3$ (O-stars).   Those values are tied to the longest-wavelength passband and are cited as approximate means owing to the acute dependence on $\ell$.  Caution is warranted with regards to the O-star estimate since emission is a concern beyond $>3.6 \mu m$, however, the stars probe key nascent regions and the young disk.

The O-star trends with Galactic longitude are consistent with the results of \citet{wh79} and \citet{pa03}.  Yet \citet{wh79} argued that local dust reputedly associated with Gould's Belt was responsible, and that is not supported by the \citet{nk80} analysis, which implies that the bulk of the extinction occurs beyond the Belt's extent.  Moreover, the present O-star results contradict theories advocating that a sizable fraction of young star forming regions should universally exhibit significantly high $R_{V}$.  The O-star and Cepheid findings likewise disfavour hypotheses that spiral arms (forefront or trailing sections) easily explain the Galactic longitude trends \citep[e.g.,][]{cr72}, as the aforementioned stellar populations may be interspersed within the arms and local structure exists (e.g., spur) possibly extending into Puppis \citep{ma09,ca15}.  A fine-tuned \textit{ad hoc} explanation may be required for that theory to explain the suite of Galactic longitude trends (O-star, classical Cepheid, and red clump).   Lastly, the recent \citet[][their Fig.~17]{sc16} results for the inner Galactic plane are broadly similar to the Galactocentric trend delineated by the classical Cepheids, however, as argued earlier the optical ($BV$) variations observed for those Cepheids are thought to stem mainly from metallicity-dependent stellar colors.

\section{Conclusion}
\label{s-conclusion}
New near-infrared observations from the $VVV$ survey were employed to help establish color-excess and total extinction ratios across the Galaxy.  To that end, Type II Cepheids and RR Lyrae variables were identified throughout the bulge and an adjacent region of the disk.  The $VVV$ observations were paired with mid-infrared observations, and used to determine the desired ratios via apparent stellar colors (\S \ref{scolors}).  The ensuing results were compared to color-excess ratios inferred from O-stars and classical Cepheids occupying the broader Milky Way.  In sum, a strict paradigm linking smaller dust grains with increasing Galactocentric distance, as inferred from red clump stars, is not supported by the results. 

The optical analysis indicates that O-stars exhibit a maximum ratio ($R_{V,BV}$) along the $\ell \sim 290 \degr$ sight-line, and minima are located toward $\ell \sim 133 \degr$ and $\ell \sim 78 \degr$ (Fig.~\ref{fig-cel}).  Conversely, the classical Cepheid optical data display extrema along the anti-centre and $\ell \sim 30 \degr$ sight-lines.  Yet the infrared colors imply a ratio ($\langle E(J-\overline{3.5 \mu m})/E(J-K_s) \rangle =1.28\pm0.03 \sigma$) that is relatively constant in comparison to optical determinations (Figs.~\ref{fig-cel} \& \ref{fig-ceg}).   The O-star color-excess ratios are particularly affected by emission beyond $3.6 \mu m$, however, the classical Cepheid, Type II Cepheid, and RR Lyrae variables may be characterized by $\langle R_{J,JK_s} \rangle =1.49\pm0.05$ (Fig.~\ref{fig-cer}), which implies $\langle A_{K_s}/A_J \rangle = 0.33\pm0.02$.   The common distance approach yielded inconsistent (lower) results relative to those inferred from stellar colors for the bulge population.  The former procedure is rather sensitive to the sample's distribution, and may be partially skewed by heavy extinction and blending (Fig.~\ref{fig-rr}). The mean optical total-to-selective extinction ratios are sensitive to $\ell$, but are approximately $\langle A_{V}/E(B-V) \rangle=\langle R_{V,BV} \rangle=3.1$ (classical Cepheids) and $\langle R_{V,BV} \rangle=3.3$ (O-stars).  The ratios are consistent with the \citet{be96} and \citet{pk12} findings, and larger than the mean determined by \citet[][$R_{V,BV}=2.40 \pm 1.05$, from high $b$ SDSS BHB stars]{ti14}.  Yet ultimately the $J,JK_s$ passband combination, and certain others \citep{ng12,na16}, present desirable advantages when establishing cosmic distances to Cepheids, which are used to constrain the Hubble constant, the Universe's age, and cosmological models \citep{fr96,ri11,ng13}.   

\subsection*{Acknowledgements}
\scriptsize{D.M. (Majaess) is grateful to the following individuals and consortia whose efforts, advice, or encouragement enabled the research: 2MASS, WISE, GLIMPSE (Spitzer), L. Berdnikov, OGLE (A. Udalski, I. Soszynski), J. Ma{\'{\i}}z-Apell{\'a}niz, N. Walborn, C. Ngeow, L. Macri, D. Balam, B. Skiff, G. Carraro, CDS (F. Ochsenbein, T. Boch, P. Fernique), arXiv, and NASA ADS.  D.M. (Minniti) is supported by FONDECYT Regular No. 1130196, the BASAL CATA Center for Astrophysics and Associated Technologies PFB-06, and the Ministry for the Economy, Development, and Tourism’s Programa Iniciativa Científica Milenio IC120009, awarded to the Millennium Institute of Astrophysics (MAS).  W.G. gratefully acknowledges financial support for this work from the BASAL Centro de Astrofisica y Tecnologias Afines (CATA) PFB-06, and from the Millennium Institute of Astrophysics (MAS) of the Iniciativa Milenio del Ministerio de Economia, Fomento y Turismo de Chile, grant IC120009.}


\begin{thebibliography}{}\setlength{\itemsep}{-1.5mm}
\bibitem[Anderson et al.(2014)]{an14} Anderson, R.~I., Ekstr{\"o}m, S., Georgy, C., et al.\ 2014, \aap, 564, A100 
\bibitem[Berdnikov et al.(1996)]{be96} Berdnikov, L.~N., 
Vozyakova, O.~V., \& Dambis, A.~K.\ 1996, Astronomy Letters, 22, 334 
\bibitem[Berdnikov et al.(2000)]{be00} Berdnikov, L.~N., Dambis, A.~K., \& Vozyakova, O.~V.\ 2000, \aaps, 143, 211 
\bibitem[Borissova et al.(2009)]{bo09} Borissova, J., Rejkuba, M., Minniti, D., Catelan, M., \& Ivanov, V.~D.\ 2009, \aap, 502, 505 
\bibitem[Caldwell \& Coulson(1985)]{cc85} Caldwell, J.~A.~R., \& Coulson, I.~M.\ 1985, \mnras, 212, 879 
\bibitem[Carraro et al.(2013)]{ca13} Carraro, G., Turner, D., Majaess, D., \& Baume, G.\ 2013, \aap, 555, A5
\bibitem[Carraro et al.(2015)]{ca15} Carraro, G., V{\'a}zquez, R.~A., Costa, E., Ahumada, J.~A., \& Giorgi, E.~E.\ 2015, \aj, 149, 12
\bibitem[Chen{\'e} et al.(2012)]{ch12} Chen{\'e}, A.-N., Borissova, J., Clarke, J.~R.~A., et al.\ 2012, \aap, 545, A54 
\bibitem[Cr{\'e}z{\'e}(1972)]{cr72} Cr{\'e}z{\'e}, M.\ 1972, \aap, 21, 85 
\bibitem[D{\'e}k{\'a}ny et al.(2015)]{de15} D{\'e}k{\'a}ny, I., Minniti, D., Majaess, D., et al.\ 2015, \apjl, 812, L29 
\bibitem[Freedman \& Madore(1996)]{fr96} Freedman, W.~L., \& Madore, B.~F.\ 1996, Clusters, Lensing, and the Future of the Universe, 88, 9 
\bibitem[Gieren et al.(2015)]{gi15} Gieren, W., Pilecki, B., Pietrzy{\'n}ski, G., et al.\ 2015, \apj, 815, 28 
\bibitem[Gontcharov(2013)]{go13} Gontcharov, G.~A.\ 2013, Astronomy Letters, 39, 83 
\bibitem[Hackwell \& Gehrz(1974)]{hg74} Hackwell, J.~A., \& Gehrz, R.~D.\ 1974, \apj, 194, 49 
\bibitem[Indebetouw et al.(2005)]{in05} Indebetouw, R., Mathis, J.~S., Babler, B.~L., et al.\ 2005, \apj, 619, 931 
\bibitem[Luck et al.(2011)]{lu11} Luck, R.~E., Andrievsky, S.~M., Kovtyukh, V.~V., Gieren, W., \& Graczyk, D.\ 2011, \aj, 142, 51 
\bibitem[Madore(1982)]{ma82} Madore, B.~F.\ 1982, \apj, 253, 575 
\bibitem[Ma{\'{\i}}z-Apell{\'a}niz et al.(2004)]{ma04} Ma{\'{\i}}z-Apell{\'a}niz, J., Walborn, N.~R., Galu{\'e}, H.~{\'A}., 
\& Wei, L.~H.\ 2004, \apjs, 151, 103 
\bibitem[Majaess et al.(2009)]{ma09} Majaess, D., Turner, D., \& Lane, D.\ 2009, \actaa, 59, 403 
\bibitem[Majaess(2010)]{ma10} Majaess, D.\ 2010, \actaa, 60, 55 
\bibitem[Majaess et al.(2012a)]{ma12a} Majaess, D., Turner, D., Gieren, W., \& Lane, D.\ 2012 (a), \apjl, 752, L10 
\bibitem[Majaess et al.(2012b)]{ma12b} Majaess, D., Turner, D., \& Gieren, W.\ 2012 (b), \pasp, 124, 1035 
\bibitem[Majaess et al.(2013)]{ma13} Majaess, D., Turner, D.~G., \& Gieren, W.\ 2013, \apj, 772, 130 
\bibitem[Macri et al.(2015)]{ma15} Macri, L.~M., Ngeow, 
C.-C., Kanbur, S.~M., Mahzooni, S., \& Smitka, M.~T.\ 2015, \aj, 149, 117 
\bibitem[Minniti et al.(2010)]{mi10} Minniti, D., Lucas, P.~W., Emerson, J.~P., et al.\ 2010, \na, 15, 433 
\bibitem[Moni Bidin et al.(2014)]{mb13} Moni Bidin, C., Majaess, D., Bonatto, C., et al.\ 2014, \aap, 561, A119 
\bibitem[Nataf et al.(2016)]{na16} Nataf, D.~M., Gonzalez, 
O.~A., Casagrande, L., et al.\ 2016, \mnras, 456, 2692 
\bibitem[Neckel et al.(1980)]{nk80} Neckel, T., Klare, G., \& Sarcander, M.\ 1980, \aaps, 42, 251 
\bibitem[Ngeow(2012)]{ng12} Ngeow, C.-C.\ 2012, \apj, 747, 50 
\bibitem[Ngeow et al.(2013)]{ng13} Ngeow, C.-C., Gieren, W., \& Klein, C.\ 2013, arXiv:1309.3481 
\bibitem[Patriarchi et al.(2003)]{pa03} Patriarchi, P., Morbidelli, L., \& Perinotto, M.\ 2003, \aap, 410, 905 
\bibitem[Pejcha \& Kochanek(2012)]{pk12} Pejcha, O., \& Kochanek, C.~S.\ 2012, \apj, 748, 107 
\bibitem[Riess et al.(2011)]{ri11} Riess, A.~G., Macri, L., Casertano, S., et al.\ 2011, \apj, 730, 119 
\bibitem[Saito et al.(2012)]{sa12} Saito, R.~K., Hempel, M., Minniti, D., et al.\ 2012, \aap, 537, A107 
\bibitem[Schlafly et al.(2016)]{sc16} Schlafly, E.~F., Meisner, A.~M., Stutz, A.~M., et al.\ 2016, \apj, 821, 78 
\bibitem[Scowcroft et al.(2011)]{sc11} Scowcroft, V., Freedman, W.~L., Madore, B.~F., et al.\ 2011, \apj, 743, 76 
\bibitem[Smith et al.(2007)]{sm07} Smith, M.~C., Wo{\'z}niak, P., Mao, S., \& Sumi, T.\ 2007, \mnras, 380, 805 
\bibitem[Sota et al.(2011)]{so11} Sota, A., Ma{\'{\i}}z Apell{\'a}niz, J., Walborn, N.~R., et al.\ 2011, \apjs, 193, 24 
\bibitem[Sota et al.(2014)]{so14} Sota, A., Ma{\'{\i}}z Apell{\'a}niz, J., Morrell, N.~I., et al.\ 2014, \apjs, 211, 10 
\bibitem[Soszy{\'n}ski et al.(2011a)]{so11a} Soszy{\'n}ski, I., Udalski, A., Pietrukowicz, P., et al.\ 2011 (a), \actaa, 61, 285 
\bibitem[Soszy{\'n}ski et al.(2011b)]{so11b} Soszy{\'n}ski, I., Dziembowski, W.~A., Udalski, A., et al.\ 2011 (b), \actaa, 61, 1 
\bibitem[Tian et al.(2014)]{ti14} Tian, H.-J., Liu, C., Hu, J.-Y., Xu, Y., \& Chen, X.-L.\ 2014, \aap, 561, A142 
\bibitem[Turner(1976)]{tu76} Turner, D.~G.\ 1976, \aj, 81,1125 
\bibitem[Turner(1994)]{tu94} Turner, D.~G.\ 1994, \jrasc, 88, 176 
\bibitem[Turner(2012)]{tu12} Turner, D.~G.\ 2012, \apss, 337, 303 
\bibitem[Whittet(1979)]{wh79} Whittet, D.~C.~B.\ 1979, \aap, 72, 370 
\bibitem[Wright et al.(2010)]{wr10} Wright, E.~L., Eisenhardt, P.~R.~M., Mainzer, A.~K., et al.\ 2010, \aj, 140, 1868 
\bibitem[Zasowski et al.(2009)]{za09} Zasowski, G., Majewski, S.~R., Indebetouw, R., et al.\ 2009, \apj, 707, 510 
\bibitem[Zechmeister \& Kurster(2009)]{zk09} Zechmeister, M., Kurster, M.\ 2009, \aap, 496, 577 
\end{thebibliography}
\end{document}